\documentclass[a4paper]{article}
\usepackage{INTERSPEECH2020}

\usepackage{cite}

\title{Multi-modal Attention for Speech Emotion Recognition}
\name{Zexu Pan$^{1,2}$, Zhaojie Luo$^{4}$, Jichen Yang$^3$, Haizhou Li$^{1,3}$}

\usepackage{xcolor}
\address{
  $^1$Institute of Data Science, NUS, Singapore\\
  $^2$Graduate School for Integrative Sciences and Engineering, NUS, Singapore\\
  $^{3}$Department of Electrical and Computer Engineering, \\National University of Singapore (NUS), Singapore\\
  $^{4}$Osaka University, Osaka, Japan}
\email{pan\_zexu@u.nus.edu, luo@irl.sys.es.osaka-u.ac.jp, \{eleyji, haizhou.li\}@nus.edu.sg}

\begin{document}

\maketitle
\begin{abstract}
Emotion represents an essential aspect of human speech that is manifested in speech prosody. Speech, visual, and textual cues are complementary in human communication. In this paper, we study a hybrid fusion method, referred to as multi-modal attention network (MMAN) to make use of visual and textual cues in speech emotion recognition. We propose a novel multi-modal attention mechanism, cLSTM-MMA, which facilitates the attention across three modalities and selectively fuse the information. cLSTM-MMA is fused with other uni-modal sub-networks in the late fusion. The experiments show that speech emotion recognition benefits significantly from visual and textual cues, and the proposed cLSTM-MMA alone is as competitive as other fusion methods in terms of accuracy, but with a much more compact network structure. The proposed hybrid network MMAN achieves state-of-the-art performance on IEMOCAP database for emotion recognition.
\end{abstract}

\begin{figure*}[ht!]
  \centering
  \includegraphics[width=\linewidth]{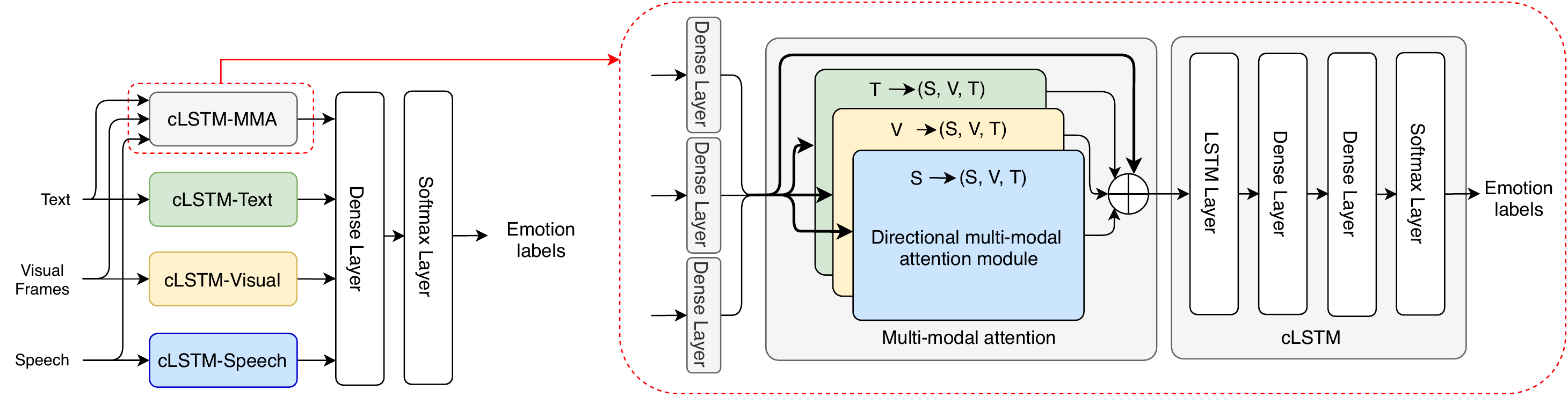}
  \caption{On the left panel is the proposed multi-modal attention network (MMAN). It consists of a multi-modal attention sub-network (cLSTM-MMA) for early fusion and three uni-modal sub-networks cLSTM-Text, cLSTM-Visual and cLSTM-Speech. The predictions of the four sub-networks are fused with a dense and a softmax layer in late fusion. The architecture of the cLSTM-MMA sub-network is shown in the red dotted box on the right panel. The symbol $\oplus$ represents concatenation and S, V, T represents speech, visual and text respectively. The cLSTM-MMA consists of three independent dense layers for uni-modal feature embeddings standardisation, multi-modal attention with three parallel directional multi-modal attention modules and finally a cLSTM with one LSTM layer inside.}
  \label{figure:network_structure}
\end{figure*}
\section{Introduction}

Emotions play an important role in speech communication\cite{sreeshakthy2016classification}. The recent advancement of artificial intelligence has equipped machines with intelligence quotient. It is equally important for machines to understand emotions, and to improve their emotional intelligence.

The fact that voice call is more informative than text messaging suggests that the affective prosody of speech delivers additional information that includes emotion. Similarly, speaking face-to-face is more effective than text messaging and voice call, which suggests that visual cues play an important role. Humans express emotion through prosody, gesture, and lexical choice. Emotion is quantized by physiological arousal and hedonic valence level \cite{barrett2006solving}, which are only partially expressed through speech. The use of specific phrases further indicates our valence level and our body language carries the remaining arousal and valence. It is found that humans rely more on multi-modalities than uni-modal \cite{shimojo2001sensory} to understand emotions.

Multi-modal speech emotion recognition has been an area of research for decades. \textit{Cho et al.} \cite{usingAudioText} used text to aid speech in the MCNN network. Similarly \textit{Hossain et al.} \cite{usingAudioVisual} and \textit{Xue et al.} \cite{xue2017bayesian} used visual cues to augment speech using SVM and Sym-cHDP networks. It is evident that emotion recognition benefits from the fusion of speech, vision and text information  \cite{poria2017review,perez2013utterance,wollmer2013youtube,poria2015deep,cambria2017benchmarking}. However, it has not been an easy task to fuse the information from different modalities. As the information coming from different modality is neither completely independent nor correlated, the fusion mechanism is expected to pick up the right information from the right modality.

Early or late fusions are the typical options in multi-modal classifier design in emotion recognition. The state-of-the-art method introduced the contextual long short-term memory block (cLSTM) and built a late fusion network (cLSTM-LF) \cite{tripathi2018multi,poria2017context}. The predictions of uni-modal models are fused to make a final prediction. It is effective at modelling modality-specific interactions but not cross-modal interactions \cite{wang2019words}.

There are also studies to explore the interaction between modalities with early fusion \cite{sebastian2019fusion,poria2016fusing,georgiou2019deep}. \textit{Sebastian et al.} \cite{sebastian2019fusion} concatenated the low-level features and passed them through a convolutional neural network. \textit{Georgiou et al.} \cite{georgiou2019deep} concatenated features from different modality at various levels and used multi-layer perceptron for emotion prediction. With early fusion, we are able to explore the interaction between raw features across modalities, that is good. However, the raw features represent different physical properties of the signals in the respective modalities. 
Therefore, the classifier network will have to learn both the feature abstraction of respective modalities, and the interaction of them at the same time, that is not easy. Furthermore, simple concatenation utilizes whatever information from the input streams that may or may not be relevant to the classification tasks. Early fusion also potentially suppresses modality-specific interactions \cite{liu2018efficient}. In general, concatenation based early fusion methods do not outperform the late fusion methods in emotion recognition \cite{poria2018multimodal,wang2019words}.

Transformer has been effective in natural language processing that features a self-attention mechanism where each input feature embedding is first projected into query, key and value embeddings \cite{vaswani2017attention}. In multi-modal situation, the query is from one modality while key and value are from another modality. The attentions between two modalities are computed by cosine similarities between the query and the key. The values are then fused based on the attention scores. The attention mechanism in Transformer is one of the effective solutions to learn cross-modality correlation \cite{tsai2019multimodal,le2019multimodal}. \textit{Tsai et al.} \cite{tsai2019multimodal} used directional pairwise cross-modal attention for sentiment analysis. 
They show positive results with two-modalities attention. In this paper, we would like to explore a mechanism for three-modalities attention for the first time. We believe that speech, visual and text modalities provide complementary evidence for emotion. Three modalities cross-modal attention allows us to take advantage of such evidence.

We propose a multi-modal attention mechanism in place of concatenation to model the correlation between three modalities in cLSTM-MMA. As cLSTM-MMA takes the multi-modal features as input, we consider it as an early fusion sub-network. It consists of three parallel directional multi-modal attention modules for multi-modal fusion. In each module, a query is first computed from a modality. It is then used to compute the cross-modal attention and the self-attention scores to find the relevant information answering to this query. The three parallel modules have distinct queries from three different modalities specifically. Thus, allowing the network to attend for different interactions based on the different queries jointly. The multi-modal attention can be easily scaled up if more than three modalities are present. To take advantage of both the late fusion and early fusion to account for modality-specific and cross-modal interactions, we propose a hybrid multi-modal attention network (MMAN) which fuses the predictions of the cLSTM-MMA and uni-modal cLSTM sub-networks for the final prediction.

The rest of the paper is organized as follows. Section 2 presents the details of the proposed multi-modal attention network. Section 3 describes the experimental setup. Section 4 reports the results and evaluations. Finally, conclusions are drawn in Section 5.

\section{Multi-modal attention network}
The proposed hybrid fusion network MMAN is shown on the left panel of Figure \ref{figure:network_structure}. We have the speech, visual and text feature embeddings of the same utterance as the input.  The MMAN consists of a cLSTM multi-modal attention sub-network (cLSTM-MMA) for early fusion, and three uni-modal sub-networks cLSTM-Speech, cLSTM-Visual and cLSTM-Text for late fusion. The outputs of the four sub-networks are fused with a dense and a softmax layer.

\subsection{Multi-modal attention}
The architecture of cLSTM-MMA sub-network is shown in the red dotted box on the right panel of Figure \ref{figure:network_structure}. The cLSTM-MMA consists of three independent dense layers for uni-modal feature embeddings standardisation, multi-modal attention with three parallel directional multi-modal attention modules and finally a cLSTM with one LSTM layer inside. 

\begin{figure}
    \centering
    \includegraphics[width=0.8\linewidth]{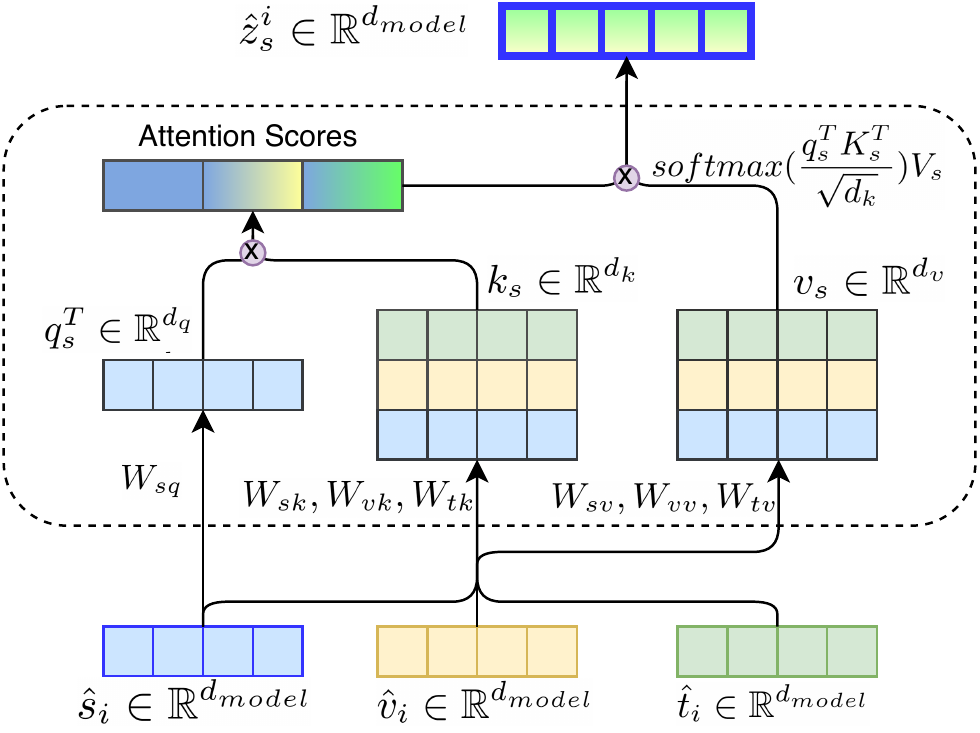}
    \caption{The details of the directional multi-modal attention module $S\xrightarrow{}(S,V,T)$ with query from speech. The inputs to this module are the uni-modal feature embeddings ($\hat{s}_i, \hat{v}_i, \hat{t}_i$) after the standardization dense layers}
    \label{figure:encoder}
\end{figure}

\subsubsection{Modality dimension standardization}
The three inputs that represent one utterance are first encoded as the feature embeddings of different dimensions. We first standardize all feature embeddings into the same dimension $d_{model}$ to facilitate the subsequent processing. 

Let's denote the dataset as $\mathcal{D} = \{s_{i},v_{i},t_{i},y_{i}\}_{i=1:M}$ where $s_{i}$, $v_{i}$, $t_{i}$ and $y_{i}$ represent the speech, visual, text feature embeddings and the emotion labels of utterance $i$. $M$ is the number of utterances in a conversation. With $s_i \in \mathbb{R}^{d_s}$, $v_i \in \mathbb{R}^{d_v}$ and $t_i \in \mathbb{R}^{d_t}$  where $d_s, d_v, d_t$ are dimensions of corresponding speech, visual and text features. By passing the original feature embeddings through the individual dense feed forward layers as shown in Figure \ref{figure:network_structure}, we standardize the outputs into the same dimension $\hat{s}_i \in \mathbb{R}^{d_{model}}$, $\hat{v}_i \in \mathbb{R}^{d_{model}}$ and $\hat{t}_i \in \mathbb{R}^{d_{model}}$.

\subsubsection{Directional multi-modal attention module}
Taking the directional multi-modal attention module with speech query for illustration. It is represented by $S\xrightarrow{}(S,V,T)$ as shown in the blue module in Figure \ref{figure:network_structure}. This module computes the directional attention from speech to visual and text as well as the self-attention of speech. The detail of this speech query module is illustrated in Figure \ref{figure:encoder}.

We use the query, key and value representation to compute the attention. We compute the query of speech $q_s$ through a learnable weights $W_{sq} \in \mathbb{R}^{d_{model} \times d_q}$ as shown in Equation \ref{equation:query}. 

\begin{equation}
\label{equation:query}
    q_s = {W_{sq}}^T\hat{s}_i
\end{equation}
where $d_q$ is the dimension of the query vector.

The keys $K_s$ and values $V_s$ are computed using learnable weights $ W_{sk},W_{vk},W_{tk} \in \mathbb{R}^{d_{model} \times d_k}, W_{sv},W_{vv},W_{tv} \in \mathbb{R}^{d_{model} \times d_v}$, where $d_k, d_v$ are dimensions of key and value vector. The computation is shown in Equation \ref{equation:key} and \ref{equation:value}. 

\begin{equation}
\label{equation:key}
    K_s = concat\{\hat{s}_i^TW_{sk}, \hat{v}_i^TW_{vk}, \hat{t}_i^TW_{tk}\}
\end{equation}
\begin{equation}
\label{equation:value}
    V_s = concat\{\hat{s}_i^TW_{sv}, \hat{v}_i^TW_{vv}, \hat{t}_i^TW_{tv}\}
\end{equation}

The cross-modal and self attention scores are computed by the dot product of the query $q_s$ and keys $K_s$. It is then used to compute the weighted sum of the values $\hat{z}_s^i$, which represents the interaction of different modalities answering to speech query. The directional multi-modal attention from speech query $D_{S\xrightarrow{}(S,V,T)}$ is given in Equation \ref{equation:speech_attention} and illustrated in Figure \ref{figure:encoder}.

\begin{equation}
    \label{equation:speech_attention}
    \begin{split}
        \hat{z}_s^i
        &= D_{S\xrightarrow{}(S,V,T)}(\hat{s}_i, \hat{v}_i, \hat{t}_i) \\
        &= softmax(\frac{q_s^TK_s^T}{\sqrt{d_k}})V_s 
    \end{split}
\end{equation}

The same computing procedure is applied to text and visual directional multi-modal attention modules except that each module has its own learnable weights computing the query to facilitate the learning of different interactions based on different directional queries. The outputs from three parallel attention modules are concatenated with a skip connection.

\subsubsection{Contextual long short-term memory block}
The output from multi-modal attention is passed through a cLSTM block with one LSTM layer as shown in Figure \ref{figure:network_structure} to capture the contextual cues between consecutive utterances in a conversation \cite{poria2017context}.

\subsection{Uni-modal sub-networks}
The cLSTM-Speech, cLSTM-Visual and cLSTM-Text sub-networks are all built using cLSTM block with two LSTM layers except that their inputs are different. Their network hyper-parameters are customized to suit different modalities. The cLSTM-MMA and three uni-modal sub-networks are separately trained. Their weights are fixed during the training of the late fusion dense layer in the MMAN.

\section{Experimental setup}

\subsection{Dataset}
The IEMOCAP dataset \cite{busso2008iemocap} is used to evaluate the proposed network. The dataset contains 10K videos split into 5 minutes of dyadic conversations for human emotion analysis. Each conversation is split into spoken utterance. Each utterance consists of corresponding transcription, speech waveform and visual frames. To align with previous works, we consider the emotion classes of angry, happy (excited), sad (frustrated) and neutral for multi-class classification but without excited and frustrated for binary sentiment classification system. The train and the test sets are disjoint for speakers. The speakers in the training set are not contained in the test set as we assume the speakers are unknown at the inference time. The details of the dataset are provided in Table~\ref{table:dataset}.

\begin{table}
    \caption{\it The number of utterances labelled happy (HPY), sad, neutral (NEU), angry (ANG), excited (EXC) and frustrated (FRU) in the training and testing set of IEMOCAP}
    \begin{tabular}{c c c c c c c} 
       \toprule
              & HPY & SAD & NEU & ANG  & EXC & FRU \\ 
       \midrule
       Train & 504 & 839 & 1324 & 933  & 742 & 1468\\ 
       Test & 144 & 245 & 384 & 170  & 299 & 381 \\ 
       \bottomrule
    \end{tabular}
    \centering
    \label{table:dataset}
    \vspace{-4mm}
\end{table}

\subsection{Uni-modal feature extraction}
We follow \textit{Poria et al.} for low level feature extraction \cite{poria2017context}. The input video of an utterance is first separated into corresponding text, video frames and speech modalities and extraction is done by using individual pre-trained networks transferred from other tasks. The feature of each utterance is extracted as a fixed-length vector for each modality.
\par
\noindent
\textbf{Speech}: OpenSMILE toolkit\cite{eyben2010opensmile} with IS13-ComParE \cite{schuller2013interspeech} is used to for feature extraction. It is performed with 30 Hz frame-rate and 100 ms sliding window. The features include Mel Frequency Cepstral Coefficient (MFCC), spectral centroid, spectral flux, beat histogram, beat sum, voice intensity, pitch, mean and root quadratic mean, etc \cite{tripathi2018multi}.
\par
\noindent
\textbf{Visual}: We use a 3D-CNN \cite{ji20123d} pre-trained from human action recognition to extract their body language. The 3D-CNN is applied to the consecutive visual frames of the speaker's upper body. It learns the relevant features of each frame and the changes among the given number of consecutive frames, which are the motion cues.
\par
\noindent
\textbf{Text}: Word2vec \cite{word2vec} is used to embed each word of an utterance's transcript into word2Vec vectors. The embedded words are concatenated, padded and standardized to a 1-dimensional vector by passing through a CNN \cite{karpathy2014large}.

\subsection{Reference baselines}
Three baselines are constructed from the state-of-the-art model \cite{poria2017context}. They are all built based on cLSTM block. The first baseline uses speech data only while the other two uses speech, visual and text data. 
\par
\noindent
\textbf{Speech-only cLSTM (cLSTM-Speech)}: The speech-only baseline receives speech features only, the speech features are passed through a cLSTM block with two LSTM layers for prediction.
\par
\noindent
\textbf{Multi-modal cLSTM with early fusion (cLSTM-EF)}: The cLSTM-EF baseline receives concatenated speech, visual and text feature embeddings as input. The concatenated features are passed through a cLSTM block with two LSTM layers for prediction.
\par
\noindent
\textbf{Multi-modal cLSTM with late fusion (cLSTM-LF)}: The cLSTM-LF baseline has a hierarchical structure. The lower level consists of three uni-modal networks cLSTM-Speech, cLSTM-Text and cLSTM-visual. At the higher level, the predictions of the three uni-modal networks are concatenated and passed through another cLSTM block for final prediction.

\section{Evaluation results}
\subsection{Comparison with baselines}
\begin{figure}[b]
  \begin{minipage}[t]{0.45\linewidth}
      \includegraphics[height=\linewidth,width=\linewidth]{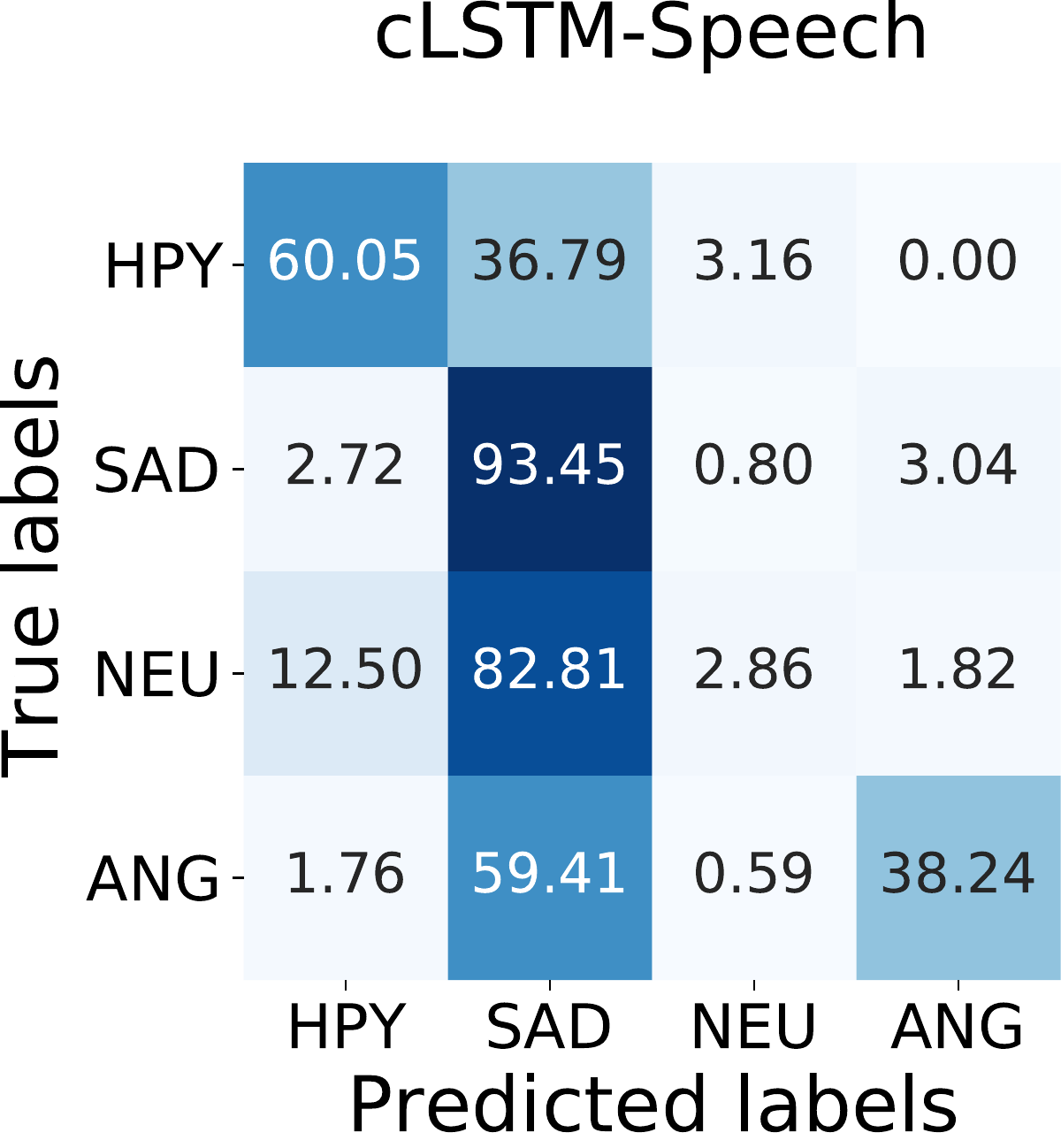}
  \end{minipage}%
      \hfill%
  \begin{minipage}[t]{0.45\linewidth}
      \includegraphics[height=\linewidth,width=\linewidth]{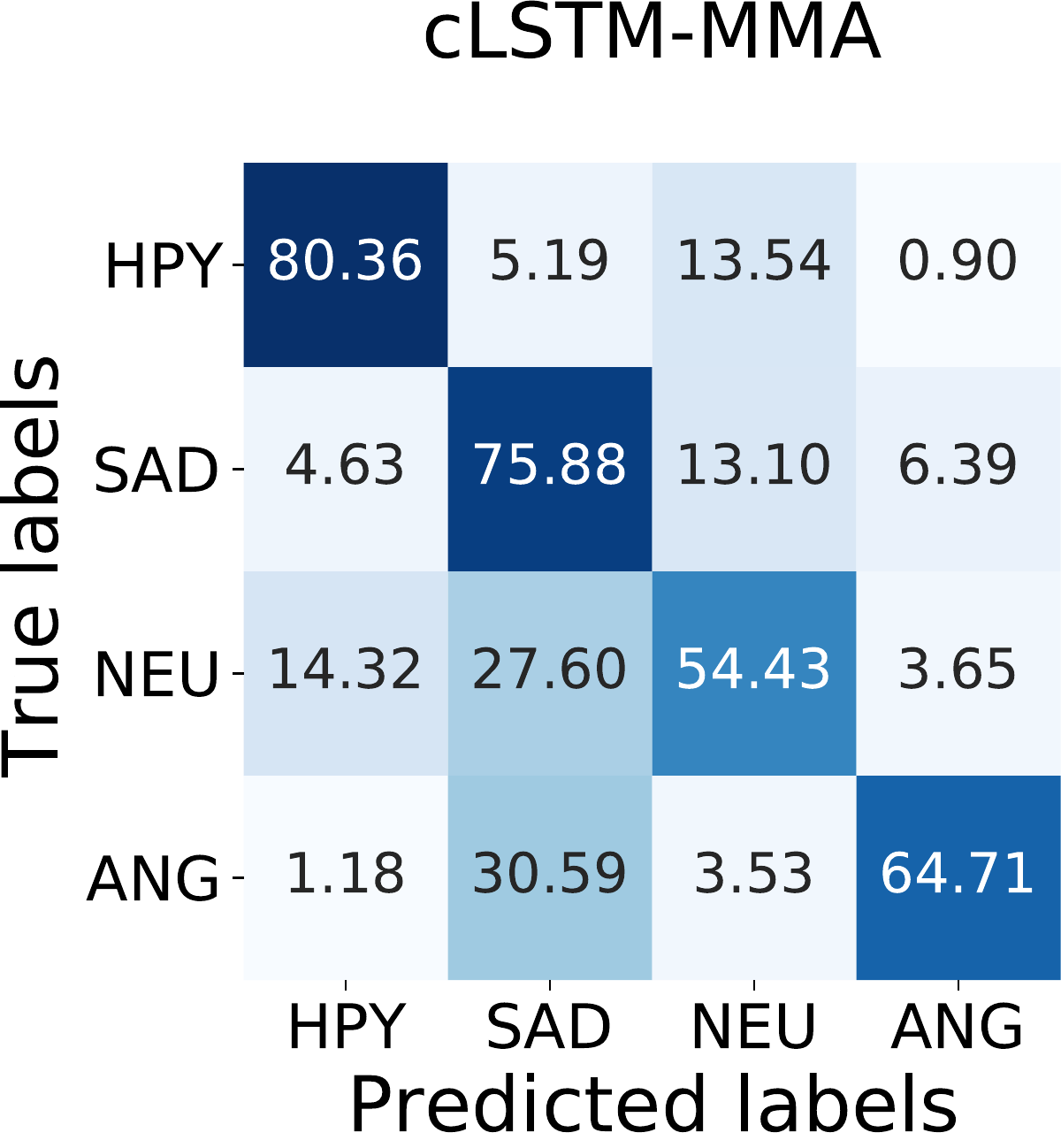}
  \end{minipage} 
      \caption{Normalised confusion matrix of the Speech-only baseline cLSTM-Speech and proposed cLSTM-MMA network. Diagonal entries represent the recall rates of each emotion.}
      \label{figure:cm}
     \vspace{-2mm}
\end{figure}

We reported the results of our proposed models and baselines of multi-class recognition systems in terms of accuracy and recall rates in Table \ref{table:accuracy} and Figure \ref{figure:cm}.

\begin{table}[t]
    \caption{Accuracy (\%) and number of network parameters of the baselines and our models in multi-class classification}
    \label{table:accuracy}
    \centering
    \begin{tabular}{c c c} 
        \toprule
         Model                 & Accuracy     & No. of Parameters (million)\\
        \midrule
         cLSTM-Speech           & 57.12             & \textbf{1.27}\\
         cLSTM-EF                & 69.75             & 2.00\\
         cLSTM-LF                & 71.78             & 4.73\\
         \midrule
         \textbf{cLSTM-MMA}         & 71.66             & \textbf{1.27}\\
         \textbf{MMAN}      & \textbf{73.94}      & 4.39\\
        \bottomrule
    \end{tabular}
\end{table}

\begin{table}[t]
  \centering
  \caption{A comparative study of binary sentiment analysis with different multi-modal implementations in terms of classification accuracy (\%)}
  \begin{tabular}{c c c c c} 
     \toprule
      Model                        & Happy & Sad & Neutral & Angry\\ 
     \midrule
     MFM \cite{DBLP:journals/corr/abs-1806-06176} &90.2 & \textbf{88.4}  &72.1   &87.5\\
     RAVEN \cite{wang2019words}      &87.3 &86.2   &69.7  & 87.3\\ 
     FMT \cite{zadeh2019factorized}   &88.8  &88.0 & 74.0 & 89.7 \\ 
     MulT \cite{le2019multimodal}    &90.7  & 86.7 & 72.4 & 87.4 \\
     \midrule
     \textbf{cLSTM-MMA}         &92.15  & 82.93 & \textbf{76.99} & \textbf{93.32} \\
     \textbf{MMAN}             &\textbf{92.68}  & 83.14 & 75.61 & 92.90 \\
     \bottomrule
  \end{tabular}
  \label{table:binary_accuracy}
  \vspace{-4mm}
\end{table}

\subsubsection{Benefits of textual and visual cues}
We first presented the recognition accuracy of the Speech-only baseline cLSTM-Speech, which obtains a recognition accuracy of 57\% which is more than 10 absolute percentage points lower than any other the multi-modal methods in Table \ref{table:accuracy}. 

We compared its confusion matrix with our cLSTM-MMA model in Figure \ref{figure:cm}, since they have similar network size. cLSTM-Speech's recall rates for neutral is very low but very high for sad as see in Figure \ref{figure:cm}. 
This unbalance phenomenon is alleviated by cLSTM-MMA, which uses multi-modal information as seen from the cLSTM-MMA confusion matrix. This shows that the visual and textual cues do complement speech's ambiguity in emotion recognition. 

\subsubsection{Multi-modal attention vs concatenation}
The cLSTM-MMA is 2\% higher than cLSTM-EF in terms of accuracy as shown in Table \ref{table:accuracy}. This means that the proposed multi-modal attention is more prevalent in computing the interaction between modalities compared to concatenation method with early fusion. Besides, the cLSTM-MMA has 40\% fewer parameters compared to the cLSTM-EF baseline. 

\subsubsection{Comparison with late fusion}
The cLSTM-MMA achieves comparable accuracy with the state-of-the-art late-fusion model cLSTM-LF with only a quarter of its' parameters as shown in Table \ref{table:accuracy}. 
The proposed hybrid MMAN network outperforms all the multi-modal networks and achieves the state-of-the-art accuracy of 73.98\% using the same amount of parameters as cLSTM-LF, suggesting that both modality-specific and cross-modal interactions are important in emotion recognition.

\begin{table}
    \centering
    \caption{A comparative study of multi-class emotion recognition with different multi-modal implementations}
    \begin{tabular}{c c} 
       \toprule
        Model                                           & Accuracy (\%)\\        
       \midrule
       \textit{Rozgic et al.}\cite{rozgic2012ensemble}        &69.4  \\
       \textit{Poria et al.}\cite{poria2018multimodal}        &71.59\\
       \textit{Tripathi et al.}\cite{tripathi2018multi}      &71.04\\    
       \midrule
       \textbf{MMAN}                                    &\textbf{73.94} \\
       \bottomrule
    \end{tabular}
    \label{table:previous_work}
     \vspace{-4mm}
\end{table}

\subsection{Comparison with previous works}
We compared the accuracies of our model with other binary sentiment classification systems using speech, visual and text in Table \ref{table:binary_accuracy}. The cLSTM-MMA has superior performance over the pairwise correlation network MulT and others (with the exception of the sad emotion). Showing that correlation between three modalities is superior then pairwise correlation. Interestingly, MMAN has similar performance with cLSTM-MMA, suggesting that modality-specific interaction may not contribute much in binary sentiment classification case.

Table \ref{table:previous_work} summaries the performance of previous multi-class emotion recognition network using speech visual and text. Our proposed MMAN achieves a state-of-the-art result of 73.94\% on dataset IEMOCAP. Also, most of the methods of the previous work that achieved comparable results are based on BLSTM which have access to future utterance information when deciding for the current utterance, thus the comparison reported in this paper is in their favour. Nevertheless, MANN outperforms all reference baselines.

\section{Conclusion}
In this work, we presented a hybrid fusion model MMAN using visual and textual cues to aid speech in emotion recognition. We proposed the multi-modal attention in early fusion which features parallel directional attention between modalities in place of concatenation. The attention mechanism enables better data association between modalities and has a significantly less amount of parameters needed. Through experiments, we showed that the multi-modal attention alone is as competitive as other fusion methods with a much more compact network. Our hybrid model achieved the state-of-the-art result on IMEOCAP dataset for emotion recognition.

\section{Acknowledgement}
This research work is partially supported by Programmatic Grant No. A1687b0033 from the Singapore Government's Research, Innovation and Enterprise 2020 plan (Advanced Manufacturing and Engineering domain), and in part by Human-Robot Interaction Phase 1 (Grant No. 192 25 00054) from the National Research Foundation, Prime Minister's Office, Singapore under the National Robotics Programme.
\bibliographystyle{IEEEtran}

\bibliography{mybib}


\end{document}